\documentclass[manuscript]{aastex631}
\usepackage{subcaption}

\begin{document}

\title{Emission-Line Ratios and Ionization Conditions of CEERS Star-Forming Galaxies with JWST/NIRSpec}

\author[0000-0003-4242-8606]{Ansh R. Gupta}
\affiliation{Steward Observatory, University of Arizona, 933 N Cherry Avenue, Tucson, AZ 85721, USA}
\correspondingauthor{Ansh R. Gupta}
\email{anshrg@utexas.edu}

\author[0000-0002-5537-8110]{Allison Kirkpatrick}
\affiliation{Department of Physics \& Astronomy, University of Kansas, Lawrence, KS 66045, USA}

\author[0000-0003-0531-5450]{Vital Fernández}
\affiliation{Michigan Institute for Data Science, University of Michigan, 500 Church Street, Ann Arbor, MI 48109, US}

\author[0000-0002-7959-8783]{Pablo Arrabal Haro}
\affiliation{NSF’s National Optical-Infrared Astronomy Research Laboratory, 950 N. Cherry Ave., Tucson, AZ 85719, USA}

\author[0000-0001-8534-7502]{Bren E. Backhaus}
\affiliation{Department of Physics \& Astronomy, University of Kansas, Lawrence, KS 66045, USA}
\affiliation{Department of Physics, 196A Auditorium Road, Unit 3046, University of Connecticut, Storrs, CT 06269, USA}

\author[0000-0001-7151-009X]{Nikko J. Cleri}
\affiliation{Department of Astronomy and Astrophysics, The Pennsylvania State University, University Park, PA 16802, USA}
\affiliation{Institute for Computational and Data Sciences, The Pennsylvania State University, University Park, PA 16802, USA}
\affiliation{Institute for Gravitation and the Cosmos, The Pennsylvania State University, University Park, PA 16802, USA}

\author[0000-0001-9440-8872]{Norman A. Grogin}
\affiliation{Space Telescope Science Institute, 3700 San Martin Drive,
Baltimore, MD 21218, USA}

\author[0000-0002-6610-2048]{Anton M. Koekemoer}
\affiliation{Space Telescope Science Institute, 3700 San Martin Drive,
Baltimore, MD 21218, USA}

\begin{abstract}
Galaxy emission-line fluxes can be analyzed to determine star formation rates (SFR) and ISM ionization. Here, we investigate rest-frame optical emission lines of 71 star-forming galaxies at redshift $0.7 < z < 7$ from the Cosmic Evolution Early Release Science (CEERS) survey using JWST/NIRSpec. We use H$\alpha$ line fluxes to measure SFRs. We combine these with HST CANDELS stellar mass estimates to determine the redshift evolution of specific SFR (sSFR) and compare our sample with the star-forming galaxy main sequence. We create [O III]$\lambda$5008/H$\beta$ versus [Ne III]$\lambda$3870/[O II]$\lambda$3728 line ratio diagrams and correlate these ratios with sSFR and the distance of each galaxy from the main sequence (excess sSFR). We find a modest correlation between the line ratios and sSFR, which is consistent with previous work analyzing similar samples. However, we find a weak correlation between the line ratios and excess sSFR. Taken together, our results suggest that sSFR is the parameter that governs ionization conditions rather than SFR or a galaxy's distance from the main sequence. These measurements reveal a rich diversity of ISM conditions and physical galaxy properties throughout cosmic time.
\end{abstract}


\section{Introduction} \label{sec:intro}
The interstellar medium is ionized by radiation from young stars. Measurements of ionization conditions thus probe physical properties of galaxies \citep[e.g.][]{2019ARA&A..57..511K}. For example, ratios between emission line fluxes of high and low ionization species reflect the hardness of the ionizing photon spectrum. Line ratio diagrams have been used to differentiate star-forming galaxies and active galactic nuclei \citep[AGN, e.g.][]{1981PASP...93....5B, 1987ApJS...63..295V, 2006MNRAS.372..961K} and probe ionization conditions of galaxies \citep[e.g.][]{2003MNRAS.346.1055K, 2024ApJ...962..195B, 2024arXiv240700157S}. The evolution of line ratios can be attributed to changes in the physical properties of galaxies over time, including star formation rates (SFR), metallicities, and AGN activity \citep[e.g.][]{2024ApJ...962...24S, 2023ApJ...954..157S}. In particular, ionization conditions echo the peak of cosmic star-formation rate density at z $\sim$ 2 and the drop in SFR that followed \citep[e.g.][]{2014ARA&A..52..415M}. Previous work has found that galaxies in a given redshift interval have specific star formation rates (sSFR $\equiv$ SFR/M${_\star}$) clustered around a central value, forming a main sequence (MS) of star-forming galaxies \citep{2007ApJ...660L..43N, 2011A&A...533A.119E}.

We investigate the ionization conditions of 71 galaxies at $0.7 < z < 7$ by analyzing the [O III]$\lambda$5008/H$\beta$ and [Ne III]$\lambda$3870/[O II]$\lambda$3728 line ratios \citep{2022ApJ...926..161B} for sources in the Extended Groth Strip (EGS) field. These ratios are derived using spectroscopic measurements from the JWST Near-Infrared Spectrograph \citep[NIRSpec,][]{2022A&A...661A..80J} as part of the Cosmic Evolution Early Release Science survey (CEERS, Finkelstein et al. in prep., PID: 1345). We examine the relationship between SFR, sSFR, and the distance of galaxies from the MS with object positions on line ratio diagrams.

In this work, we use a $\Lambda$CDM cosmology with $\Omega_\Lambda = 0.7$, $\Omega_m = 0.3$, and $H_0 = 70$ km s$^{-1}$ Mpc$^{-1}$.

\section{Analysis}
The EGS field is covered by JWST/NIRSpec medium-resolution grating and prism spectroscopy as part of CEERS \citep[e.g.][]{2023ApJ...949L..25F, 2023ApJ...951L..22A}. Data reduction is discussed in Arrabal Haro et al. (in prep.) Redshifts and line flux measurements for sources were made using the LIne MEasuring library \citep{2023MNRAS.520.3576F} and results were visually inspected to verify the validity of the fits. SFRs were derived from H$\alpha$ \citep{2012ARA&A..50..531K}.

This sample also has HST ACS and WFC3 photometry. This coverage was obtained as part of the CANDELS program \citep{2011ApJS..197...35G, 2011ApJS..197...36K}. Stellar mass estimates were determined through photometric SED fitting \citep{2017ApJS..229...32S}.

\section{Results}
\begin{figure*}[ht!]
    \centering
    \begin{subfigure}{\textwidth}
    \includegraphics[width=0.7\textwidth]{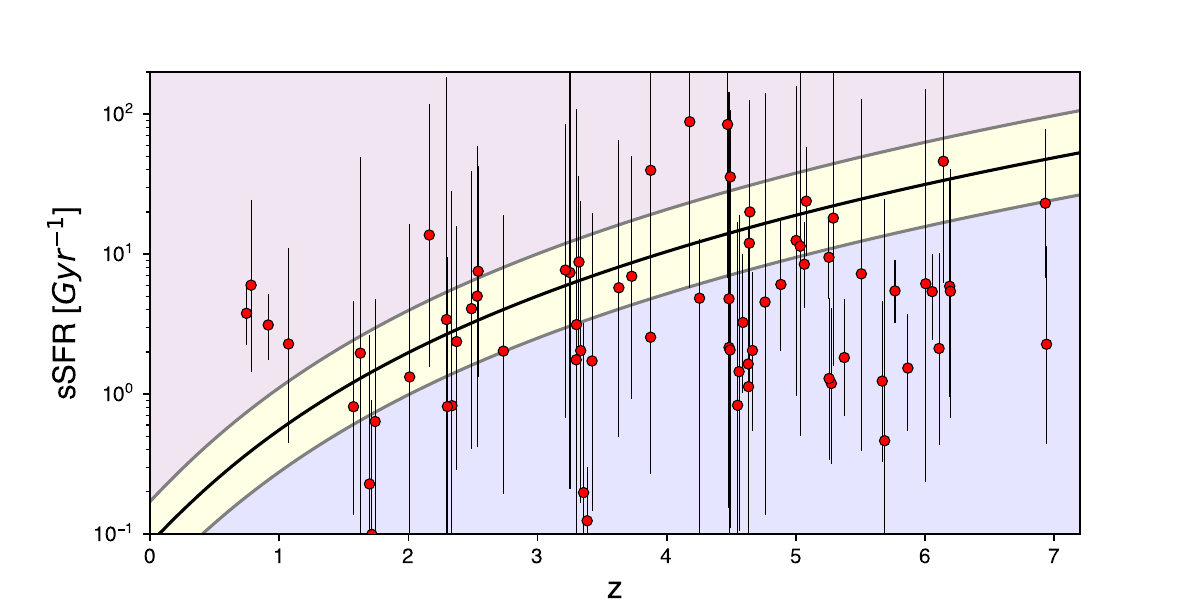}
    \centering
    \subcaption{The star-forming galaxy main sequence defined by \citet{2011A&A...533A.119E}. We neglect dust attenuation, making these sSFR measurements lower limits. The sSFR error is dominated by uncertainty in stellar mass. Populations above and below the MS (yellow stripe) reveal starbusting and quenching galaxies.}
    \label{fig:a}
    \end{subfigure}

    \begin{subfigure}{\textwidth}
    \includegraphics[width=0.8\textwidth]{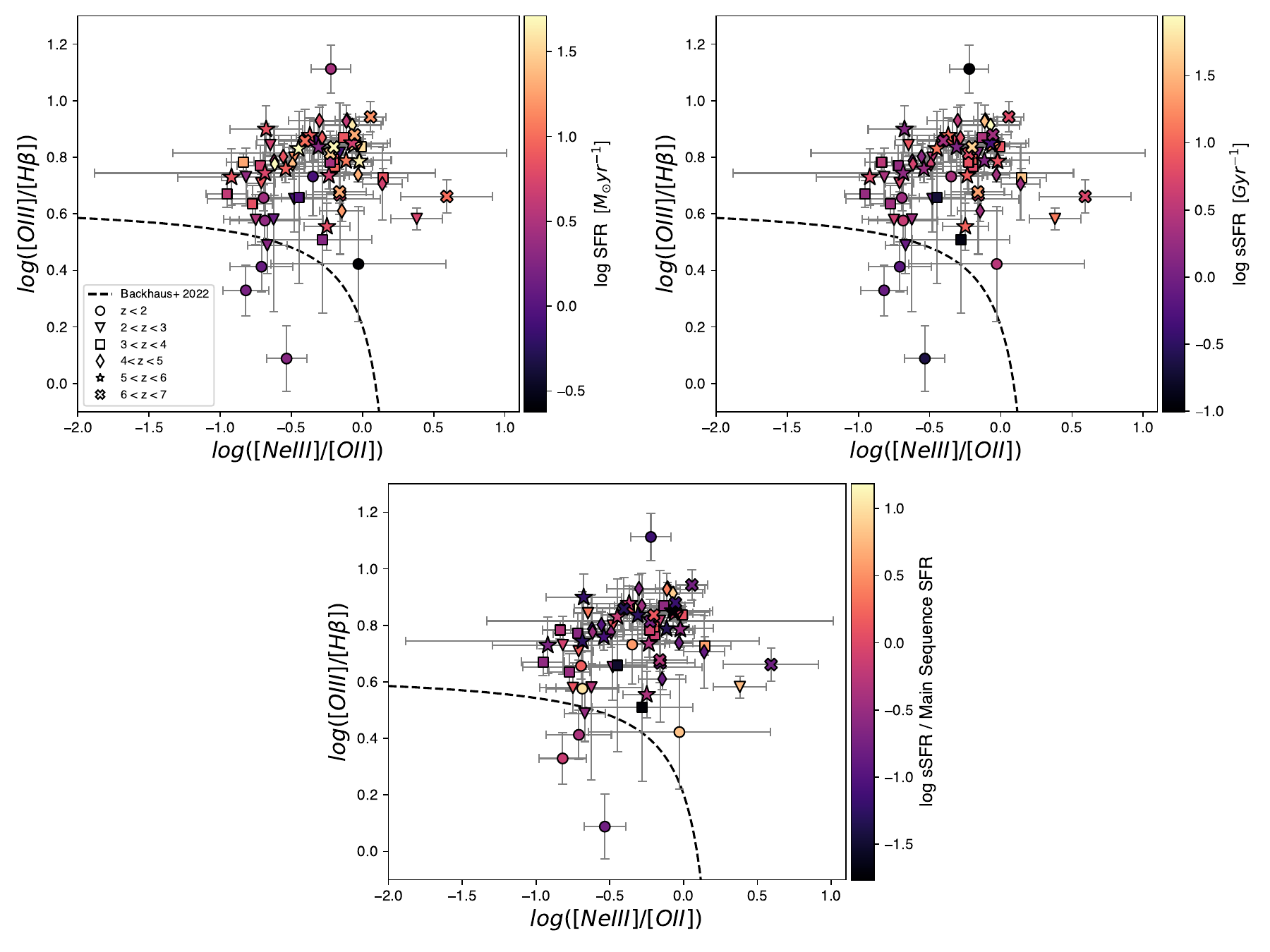}
    \centering
    \subcaption{[O III]$\lambda$5008/H$\beta$ versus [Ne III]$\lambda$3870/[O II]$\lambda$3728 line ratio diagrams, shaded by SFR, sSFR, and excess sSFR. The dashed black line is the empirical division at $z \sim 1.5$ between star-forming galaxies (below) and AGN (above) \citep{2022ApJ...926..161B}. The ionization of galaxies appears most strongly correlated with sSFR.}
    \label{fig:b}
    \end{subfigure}
    \caption{Star formation relative to the star-forming galaxy main sequence and correlation between emission line flux ratios and SFR, sSFR, and excess sSFR of CEERS star-forming galaxies.}
\end{figure*}

\autoref{fig:a} shows positions of the selected sources on the MS \citep{2011A&A...533A.119E}. The scatter about the MS reflects the underlying diversity of objects in the observed field \citep[e.g.][]{2023ApJ...955...54S}.

\autoref{fig:b} displays the modest redshift evolution of the [O III]$\lambda$5008/H$\beta$ and [Ne III]$\lambda$3870/[O II]$\lambda$3728 line ratios, consistent with the results of \citet{2023ApJ...945...35T}. Nearly all sources lie above the empirical $z \sim 1.5$ line dividing AGN and star-forming galaxies \citep{2022ApJ...926..161B}, emphasizing the need to recalibrate such metrics at higher redshifts \citep[e.g.][]{2023ApJ...953...10C, 2023ApJ...958..141L}. Alternatively, some sources may have a partial contribution from AGN emission.

SFR does appear to correlate with galaxy ionization, consistent with results from similar samples \citep[e.g.][]{2024ApJ...962..195B, 2022ApJ...937...22P}. However, sSFR appears more strongly correlated. Surprisingly, excess sSFR does not appear to be strongly correlated with either of the line ratios. This comparison suggests that sSFR can better account for the ionization of gas in a galaxy than the total amount of radiation present or its distance from the MS.

CEERS 2919, the object with the highest [O III]$\lambda$5008/H$\beta$ ratio, is confirmed as an AGN via the detection of a broad H$\alpha$ component and an associated X-ray emission source \citep{2024arXiv240815615M} in the Chandra AEGIS-XD survey \citep{2015ApJS..220...10N}, explaining its high gas ionization despite its low sSFR.

\section{Conclusions}
We analyze the redshift evolution of the [O III]$\lambda$5008/H$\beta$ and [Ne III]$\lambda$3870/[O II]$\lambda$3728 line ratios of 71 star-forming galaxies using JWST/NIRSpec observations from the CEERS survey. The measured trends appear consistent with previous works which indicate higher ionization and lower metallicity at greater redshifts. Our results suggest that sSFR is more responsible for the ionization conditions of galaxies than SFR or distance from the MS.

\begin{acknowledgements}
A.G. was supported by the National Science Foundation through grant $\#2149897$. V. F. research project is supported by the Eric and Wendy Schmidt AI in Science Postdoctoral Fellowship, a Schmidt Futures Program.
\end{acknowledgements}

\bibliography{sample631}{}
\bibliographystyle{aasjournal}

\end{document}